\documentclass[twocolumn,amsmath,amssymb,showpacs,prl]{revtex4-1}

\usepackage[usenames,dvips]{color}
\usepackage{color}
\usepackage{graphicx}
\usepackage{dcolumn}
\usepackage{bm}
\usepackage{soul}

\newcommand{\be}[1]{\begin{equation}\label{eq:#1}}
\newcommand{\ee}{\end{equation}}
\newcommand{\bea}{\begin{eqnarray}}
\newcommand{\eea}{\end{eqnarray}}

\newcommand{\bt}{\textbf}

\newcommand{\phd}{\phantom{\dag}}
\newcommand{\ph}{\phantom{.}}

\newcommand{\up}{^{\phd}}
\newcommand{\noi}{\noindent}
\newcommand{\no}{\nonumber}

\begin{document}
\title{Circular-polarization-sensitive metamaterial based on triple quantum-dot molecules}
\author{Panagiotis Kotetes$^{1,2}$}
\email{panagiotis.kotetes@kit.edu}
\author{Pei-Qing Jin$^3$}
\email{pqjin@shmtu.edu.cn}
\author{Michael Marthaler$^1$}
\author{Gerd Sch\"on$^{1,2}$}
\affiliation{$^1$Institut f\"ur Theoretische Festk\"orperphysik, Karlsruhe Institute of Technology, 76128 Karlsruhe, Germany}
\affiliation{$^2$DFG Center for Functional Nanostructures (CFN), Karlsruhe Institute of Technology, 76128 Karlsruhe, Germany}
\affiliation{$^3$Institute of Logistics Engineering, Shanghai Maritime University, Shanghai 201306, China}

\begin{abstract}
We propose a new type of chiral metamaterial based on an ensemble of artificial molecules formed by three identical quantum-dots in a triangular arrangement. A static magnetic
field oriented perpendicular to the plane breaks mirror symmetry, rendering the molecules sensitive to the circular polarization of light. By varying the orientation and
magnitude of the magnetic field one can control the polarization and frequency of the emission spectrum. We identify a threshold frequency $\Omega$, above which we find strong
birefringence. In addition, Kerr rotation and circular-polarized lasing action can be implemented. We investigate the single-molecule lasing properties for different
energy-level arrangements and demonstrate the possibility of circular polarization conversion. Finally, we analyze the effect of weak stray electric fields or deviations from
the equilateral triangular geometry.

\end{abstract}



\maketitle

Light with circular polarization (CP) has a broad range of applications, e.g., for spintronics devices requiring coherent spin control \cite{Spintronics} or optical
communication with spin-based information processing \cite{OptComm}. Efficient gene\-ra\-tion, manipulation, and detection of CP requires materials with broken
mirror symmetry, either due to the structure or the violation of time-reversal symmetry (${\cal T}$). Examples of CP sensitive materials include chiral semiconducting
nanostructures employed as CP light emitters \cite{CPLE}, photonic metamaterials involving gold helices \cite{Metam}, and spin lasers based on III-V type semiconducting
quantum-dots (QDs)\cite{SpinLaser}. However, tuning the CP emis\-sion charac\-te\-ri\-stics, which would open perspectives for further applications, remains challen\-ging.
E.g., the structurally chiral ma\-te\-rials can not be manipulated due to their built-in han\-ded\-ness, while the functionality of spin-lasers is li\-mi\-ted by the
spin-injection efficiency \cite{SpinLaser}. A step towards this goal has been made very recently with the realization of an electrically switchable CP light emitter
\cite{ChiralLET}. 

Further progress along this direction can be made u\-sing QD systems, which combine the versatility of nanoelectronics with the advantages of structurally chiral
metamaterials. Specifically we suggest to use ensembles of QD molecules, each consisting of three identical QDs arranged in a triangular fashion. Such QD molecules recently
became accessible \cite{GaudreauFirst,Schroer,Ihn,Rogge,Amaha1,Laird,Seo,TQDReview}. An external magnetic field can influence the QD molecules by coupling to both spin and
orbital degrees of freedom. Linear molecules are currently investigated for applications based on Zeeman-splitting, for quantum enginee\-ring based on the exchange-qubit
protocol \cite{Laird,ExchangeQubit}, as well as for spin-blockade \cite{SpinBlockade} effects. Due to their topo\-lo\-gy, the triangular quantum-dot (TQD) molecules display
additional orbital effects. When a magnetic field is applied with orientation perpendicular to the plane, the TQD electronic wavefunctions acquire chira\-li\-ty. So far, the
influence of the magnetic flux on TQD molecules has been considered in the context of spin-chirality-coded qubits \cite{ChiralityI,ChiralityII}, Aharonov-Bohm oscillations
\cite{AB}, and transport properties \cite{Delgado,Flux}.

\begin{figure}[t]
\centering
\includegraphics[width=0.85\columnwidth]{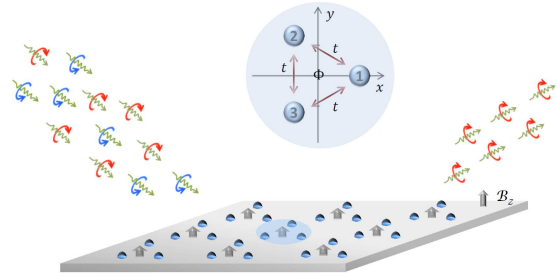}
\caption{Triangularly arranged triple quantum-dot molecules as building blocks of a chiral metamaterial. In a perpendicu\-lar magnetic field (${\cal B}_z$) the device 
becomes sensitive to the circular pola\-ri\-za\-tion of light. Controllable Kerr rotation and circularly po\-la\-ri\-zed lasing action can be realized. By varying the 
magnetic flux $\Phi$ through each molecule one can control the circular polarization and frequency of the emitted light.}
\label{Fig:fig1}
\end{figure}

In this work we address a yet unexplored aspect of TQD molecules, namely their flux-tunable CP sen\-si\-ti\-vi\-ty, which is fundamental for the operation of the chiral
metamaterial that we propose. Each TQD molecule exhibits CP birefringence and thus can serve as a Kerr rotator. CP birefringence becomes strong above a thre\-shold frequency
$\Omega$, where the TQD molecule becomes completely transparent to one of the two CPs. Furthermore, we propose the TQD molecule as an active medium for CP lasing action. A
broadband pumping field can create a popu\-lation inversion, and light with controlled circular pola\-ri\-zation is emitted. The energy level hierarchy of the TQD molecule,
which can be manipulated via the magnetic flux, determines which CP dominates for each molecular transition. This allows for a tunable switching of the CP of the emitted light
by experimentally accessible magnetic fields.

For our description, we consider three \textit{identical} QDs located at $\bm{R}_1=(a,0)$ and $\bm{R}_{2/3}=(-a/2,\pm \sqrt{3}\,a/2)$, as in Fig.~\ref{Fig:fig1}. We assume a
single energy level $\epsilon$ for each QD, as also interdot tunneling of strength $t$. We consider a single spin species and work with a dot basis
$\{\left|1\right>,\left|2\right>,\left|3\right>\}$. This is justified since we focus on single-electron effects leading to a spin-independent emission spectrum, while
spin-mixing terms, such as spin-orbit coupling, are weak. In the absence of the magnetic field the system possesses a $C_{3v}$ point group symmetry, generated by a
$\widehat{C}_3$ ($2\pi/3$) counter-clockwise rotation of the system about the $\bm{\hat{z}}$ axis and a mirror operation $\widehat{\sigma}_v$ ($y\leftrightarrow-y$). The $C_3$
symmetry suggests to switch to the more ap\-pro\-priate chiral basis
\bea
\left|\lambda\right>=\frac{1}{\sqrt{3}}\left(1, \, e^{-\lambda 2\pi i/3}, \, e^{\lambda 2\pi i/3}\right)^{\rm T}\phantom{-}{\rm with}\phantom{-}\lambda=0,\pm\,,\quad
\eea

\noindent where the superscript $\rm T$ denotes matrix transposition. The chiral basis states satisfy $\widehat{C}_3\left|\lambda\right>=e^{i \lambda
2\pi/3}\left|\lambda\right>$ with $\lambda=0,\pm$. Mirror symmetry implies that $\left|\pm\right>$ are de\-ge\-ne\-ra\-te, additionally reflecting the preservation of 
${\cal T}$.

In the presence of a perpendicular magnetic field $\bm{{\cal B}}={\cal B}_z\bm{\hat{z}}$, the Hamiltonian in the dot basis $\{\left|1\right>,\left|2\right>,\left|3\right>\}$
reads
\bea
\widehat{{\cal H}}_{\rm TQD}=\left(
\begin{array}{ccc}
\epsilon& -\,t\,e^{\,i\phi}&-\,t\,e^{-i\phi}\\
-\,t\,e^{-i\phi}&\epsilon& -\,t\,e^{\,i\phi}\\
-\,t\,e^{\,i\phi}&-\,t\,e^{-i\phi}&\epsilon
\end{array}
\right)\,,\label{Eq:Hamiltonian}
\eea

\noi where $\phi=2\pi\nu/3$, and $\nu=\Phi/\Phi_0$ denotes the normalized flux ($\Phi_0=h/e$) piercing the triangular area $A=3\sqrt{3}\,a^2/4$ of the TQD molecule. The
Hamiltonian is diagonal in the chiral basis with eigenenergies, $E_{\lambda}=\epsilon-2\,t\cos[2\pi(\nu-\lambda)/3]$. For half-integer multiples of three flux quanta, mirror
and ${\cal T}$ symmetries are restored, leading to the aforementioned de\-ge\-ne\-racy between the $\left|\pm\right>$ states. For the remaining half-integer multiples of
$\Phi_0$, additional degeneracies appear between the $\left|0\right>$ and the $\left|\pm\right>$ le\-vels. If the dots are considered as ideal zero-dimensional objects,
$\epsilon$ and $t$ are flux-independent, rendering the energy spectrum periodic in the flux with a period of three flux quanta. On the other hand, for finite-size dots,
$\epsilon$ and $t$ depend on the flux due to the orbital effects of the applied magnetic field. The latter dependence can be obtained via a microscopic continuum model
\cite{Burkard99,Sup}. As a result, the flux periodicity is broken, while the degeneracies persist.

\begin{figure}[t]
\centering
\includegraphics[scale=0.55]{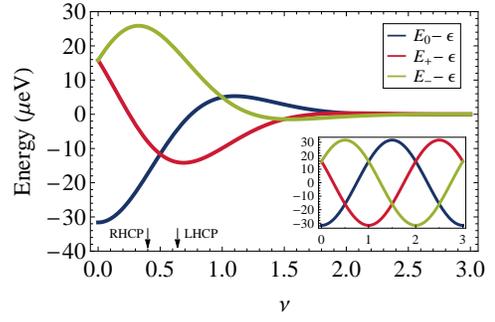}
\caption{Eigenenergies, $E_{\lambda}-\epsilon$, of a TQD molecule versus the normalized flux $\nu$ as determined from a microscopic analysis \cite{Sup}. We choose
$\hbar\,\omega_0 = 460 \,\mu\rm{eV} $ with Bohr radius $a_{\rm B} =\,50\,\rm{nm}$ and interdot distance $200 \rm{nm}$. For comparison the inset displays the periodic spectrum
obtained when $\epsilon$ and $t$ are assumed to be flux-independent. To illustrate the CP sen\-si\-ti\-vi\-ty we mark two specific values, $\nu = 0.40$ and $\nu = 0.64$, where
the transitions between the states $|+\rangle$ and $|0\rangle$ lead to photons with the same frequencies but with right- and left-handed CP (RHCP and LHCP), respectively.}
\label{Fig:fig2}
\end{figure}

The flux dependent eigenenergies of the TQD molecule, measured relative to $\epsilon$, are depicted in Fig.~\ref{Fig:fig2}. Apart from the dege\-ne\-ra\-cies occuring at
half-integer va\-lues of $\nu$, we observe that the energy differences, $E_{\lambda}-\epsilon$, decrease with increasing flux as a result of the field dependence of the
interdot tunneling strength $t$. Essentially the latter sets the range of the Bohr frequencies $\omega_{\lambda, \lambda'} \equiv (E_\lambda-E_{\lambda'})/\hbar$ and thus
the frequencies of the emitted or absorbed light. Furthermore, the level hie\-rarchy determines the selection of CP for each mo\-le\-cu\-lar transition. Its tunability, via
the flux dependence of the spectrum, is crucial for the CP of the emission spectrum. Note finally that $E_{\lambda}$ exhibit an increasing trend with the enclosed flux due to
the diamagnetic shift of $\epsilon$ \cite{Sup,FDStates,Delgado}.

The CP-sensitivity of the TQD molecule becomes mani\-fest in its dipole coupling to light, which is described by $\widehat{{\cal H}}_{\rm dip}=-\widehat{\bm{P}}\cdot
\widehat{\bm{{\cal E}}}$, with polarization operator $\widehat{\bm{P}}=-e\sum_{n=1,2,3}\left|n\right>\bm{R}_n\left< n\right|$. In the chiral basis
\bea
\widehat{P}_x=-\frac{ea}{2}\left(\begin{array}{ccc}0&1&1\\1&0&1\\1&1&0\end{array}\right)\,,\phd
\widehat{P}_y=-\frac{ea}{2}\left(\begin{array}{ccc}0&-i&i\\i&0&-i\\-i&i&0\end{array}\right)\,,\quad
\eea

\noindent with elementary charge $e>0$. It is more convenient to transfer to the CP basis, i.e., $\widehat{{\cal H}}_{\rm dip}=-\widehat{P}_+\widehat{{\cal
E}}_--\widehat{P}_-\widehat{{\cal E}}_+$, with $\widehat{P}_{\pm}=(\widehat{P}_x\pm i\widehat{P}_y)/\sqrt{2}$ and $\widehat{{\cal E}}_{\pm}=(\widehat{{\cal E}}_x\pm
i\widehat{{\cal E}}_y)/\sqrt{2}={\cal E}(\hat{a}_{\pm}+\hat{a}_{\mp}^{\dag})$, where $+/-$ corresponds to right/left-handed CP (RHCP/LHCP). We introduced the CP photonic
operators $\hat{a}_{\pm}=(\hat{a}_x\pm i\hat{a}_y)/\sqrt{2}$. The prefactor ${\cal E} = \sqrt{\hbar\,\omega/(2\varepsilon v)}$ depends on the dielectric constant $\varepsilon$
and the effective mode vo\-lu\-me $v$. With the use of $\hat{a}_{\pm}$, we obtain 
\bea
\widehat{{\cal H}}_{\rm dip}=\hbar g
\left(\begin{array}{ccc}0&1&0\\0&0&1\\1&0&0\end{array}\right)\left(\hat{a}_{+}^{\dag}+\hat{a}_{-}\up\right)+{\rm H.c.}\,,\label{eq:coupling}
\eea

\noindent with coupling strength $g=e{\cal E}a/\sqrt{2}\hbar$, that depends on the interdot distance and the operational frequency domain.

For a given energy-level hierarchy, each molecular transition becomes CP-filtered. For illustration, we focus on the transition between the states $\left|0\right>$ and
$\left|+\right>$, for which the dipole interaction reads
\bea\widehat{{\cal H}}_{\rm dip}^{(0,+)}=
\hbar g\big(\hat{a}_+^{\dag}+\hat{a}_-\big)\left|0\right>\left<+\right|+{\rm H.c.}\,.\eea

\begin{figure}[t]
\centering
\includegraphics[width=0.74\columnwidth]{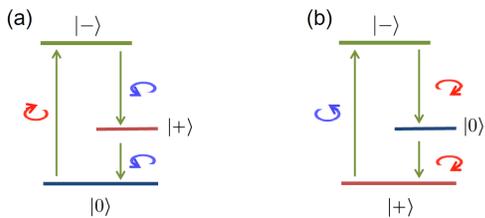}
\caption{Illustration of the CP-sensitivity of the TQD molecule for two values of the applied magnetic field, corresponding to $\nu=0.40$ (a) and $\nu=0.64$ (b). Depending on
the energy-level arrangement, every molecular transition leads to emitted or absorbed light of a particular CP. Here blue/red denotes right/left-handed CP. Within the
rotating-wave approximation, if the lowest energy level is occupied, the absorbed and emitted light always have opposite CPs (blue/red), thus resulting in polari\-za\-tion
conversion.}
\label{Fig:Dipole}
\end{figure}

\noi When the state $\left|+\right>$ is occupied and lies higher in ener\-gy than $\left|0\right>$, as in Fig.~\ref{Fig:Dipole}(a), the process
$\left|+\right>\rightarrow\left|0\right>$ is accompanied by the emission of a right-handed photon ($\propto \hat{a}_+^{\dag}$) with probability ${\rm p}_+=1$. In
con\-trast, the emission of a left-handed photon is forbidden as ${\rm p}_-=0$, while the absorption of a left-handed photon ($\propto \hat{a}_-$) is possible.
However, the latter is strongly suppressed and ignored in the rotating wave approximation \cite{PI}, since it involves a fast oscillation. When the ener\-gy level arrangement
for the levels is inverted, now with the $\left|0\right>$ state occupied as in Fig.~\ref{Fig:Dipole}(b), left-handed photons are emitted. Consequently, by controlling the
energy level arrangement via the magnetic flux, one can select the polarization of photons which are emitted by the TQD molecule. Based on these properties, the TQD molecule
can be used for CP-\textit{conversion}, i.e., the CP of the photons which are absorbed when exciting the molecule from the lowest to the highest le\-vel, is opposite to the CP
of the photons emitted when the molecule is de-excited.

\begin{figure}[t]
\centering
\includegraphics[width=0.75\columnwidth]{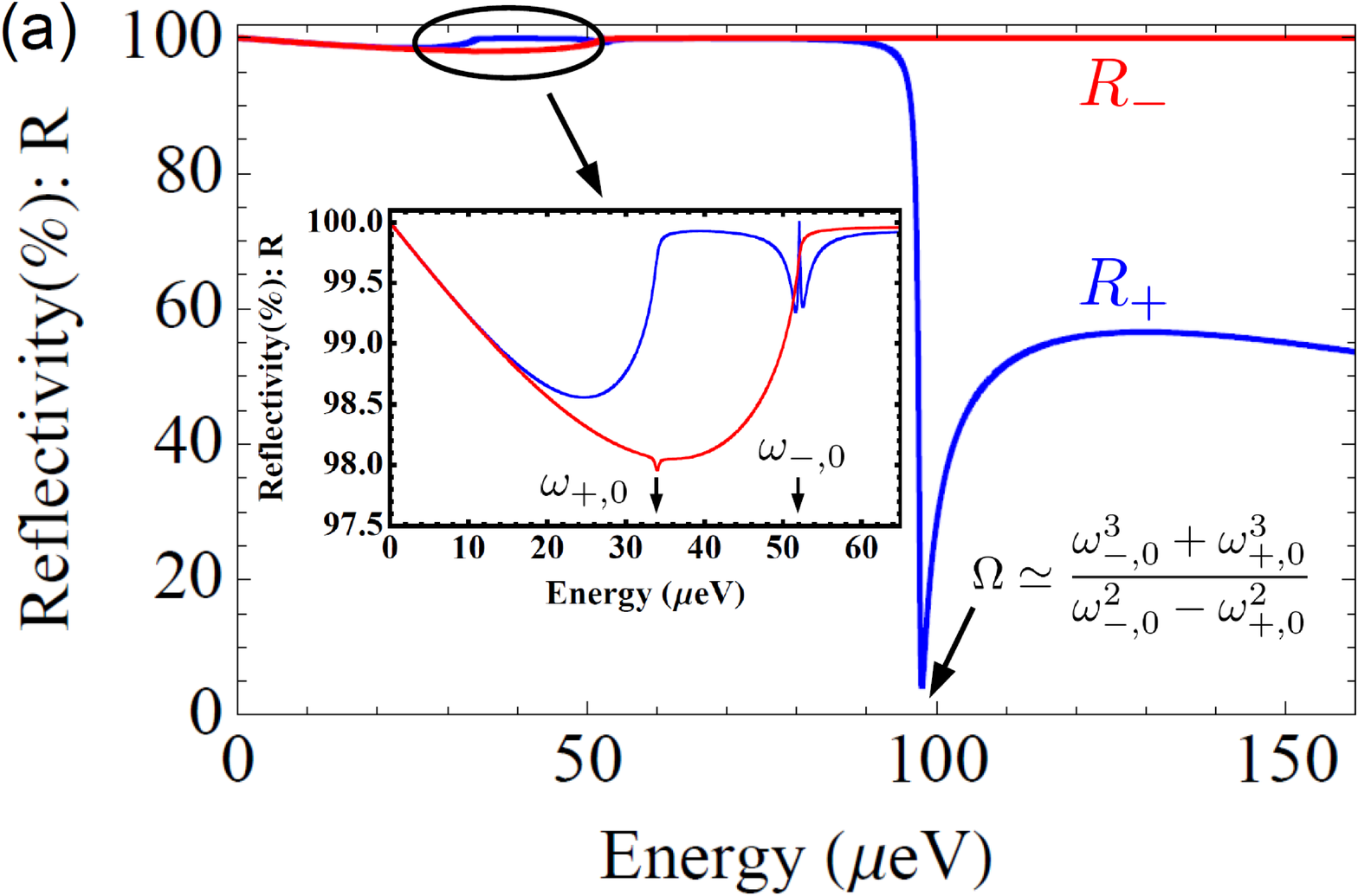}\vspace{0.07in}\\
\includegraphics[width=0.42\columnwidth]{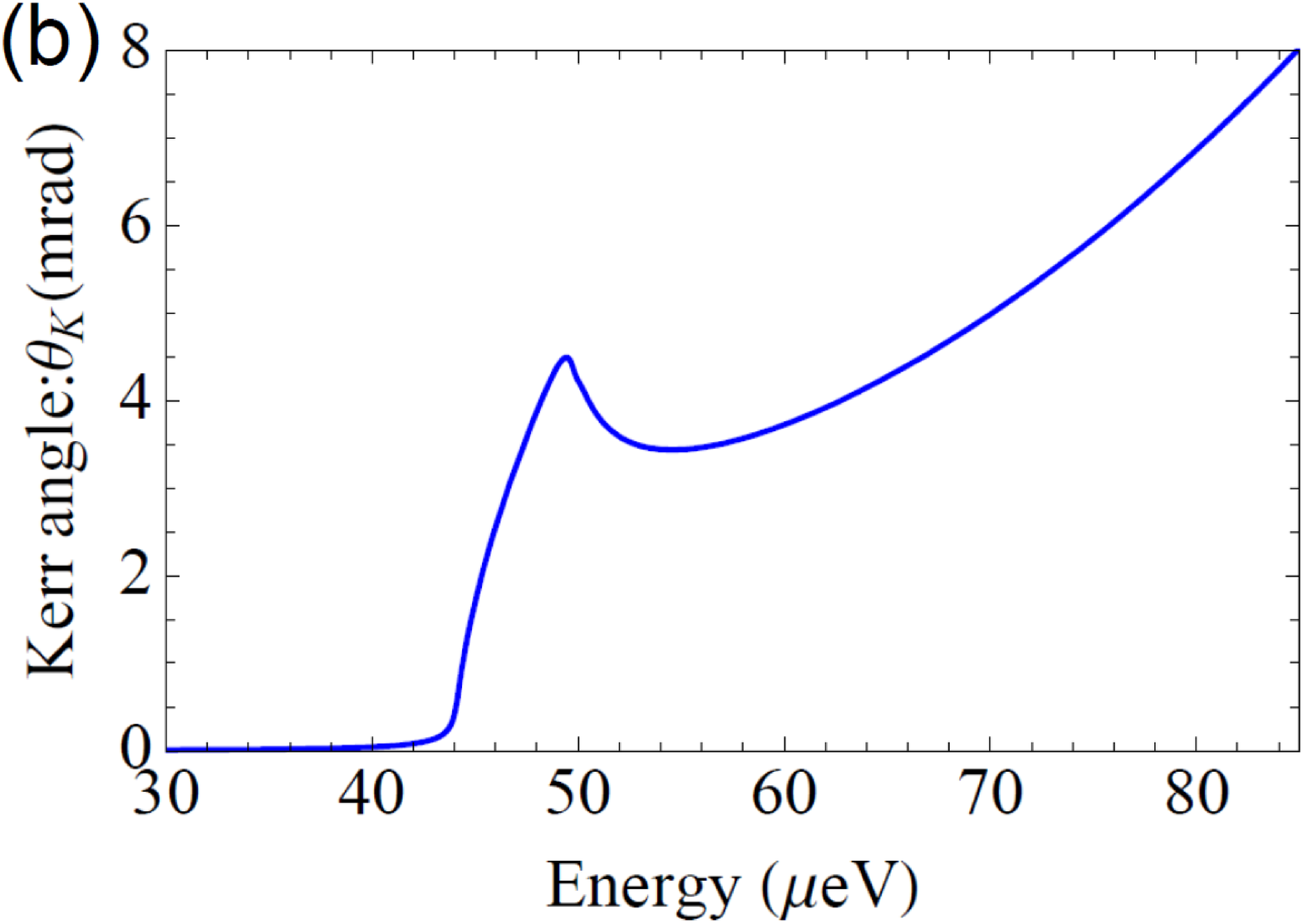}\hspace{0.3in}
\includegraphics[width=0.27\columnwidth]{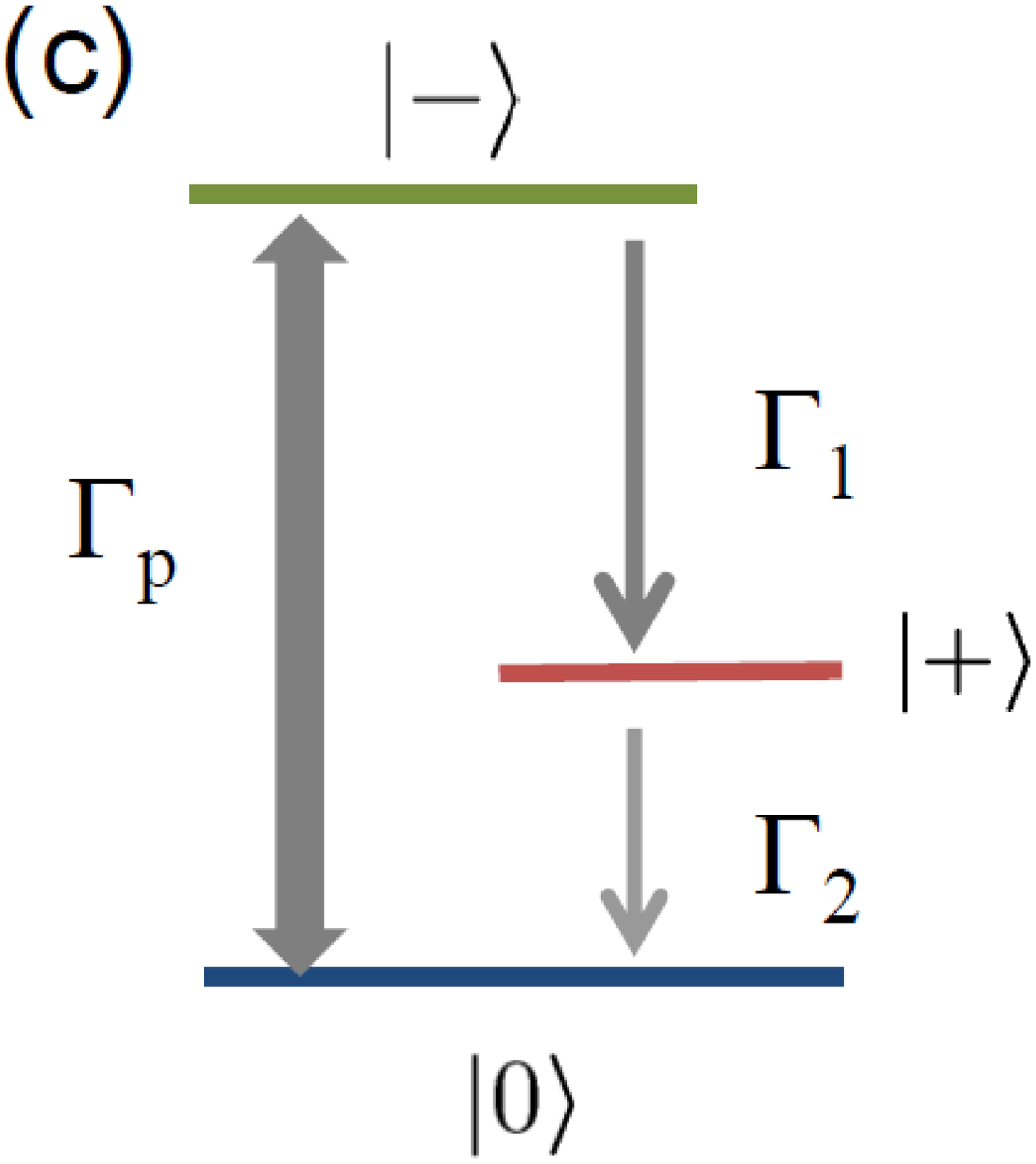}
\caption{(a) Reflectivities for RHCP (blue) and LHCP (red) at $\nu=1/6$. For photon energies close to the two TQD Bohr ener\-gies, $\hbar\omega_{+/-,0}\simeq 34/52{\rm \mu
eV}$, we find small dips, but still both CPs are almost perfectly reflected (see Inset). For a particular photon frequency, $\Omega$, we find complete transparency for the
RHCP, i.e. $R_+=0$, while the LHCP is perfectly reflected. $\Omega$ constitutes a threshold frequency, above which strong CP birefringence occurs, since a significant
difference between the reflectivities of the two CPs persists, with $R_-\geq 2R_+$. (b) The Kerr angle becomes sizeable for $\omega>\omega_{+,0}$, shows a sharp peak for
$\omega\simeq\omega_{-,0}$ and remains large beyond. Here we choose $\nu=1/20$ which yields $\hbar\omega_{+/-,0}\simeq 44/50{\rm \mu eV}$. (c) Pumping scheme for a
CP-tunable laser. A broadband pumping field induces transitions between states $\left|-\right>$ and $\left|0\right>$ (with equal excitation and relaxation rates $\Gamma_{\rm
p}$), while the lasing transition involves the states $|+\rangle$ and $|0\rangle$. A population inversion occurs when the relaxation from state $\left|-\right>$ (with rate
$\Gamma_1$) is stronger than that from state $\left|+\right>$ (with rate $\Gamma_2$).}
\label{Fig:KerrPumping}
\end{figure}

The properties of the TQD molecules open new perspectives in CP-sensitive applications. Three examples are described below: (i) CP-birefringence and (ii) tu\-na\-ble Kerr
rotation, both considered when in the presence of a static magnetic field ${\cal B}_z$ a classical electromagnetic wave ($\bm{{\cal E}}(z,t)=\bm{{\cal E}}e^{i(qz-\omega t)}$)
is normally incident to the TQD $xy$-plane, and (iii) lasing action with a TQD-based metamaterial as an active medium, when the latter is coupled to a cavity sup\-por\-ting
CP, e.g., a vertical ca\-vi\-ty used in a spin-laser \cite{SpinLaser}. The measurement of the CP-dependent reflec\-ti\-vi\-ties, $R_{\pm}$, and the resulting Kerr angle,
$\theta_K$, can serve as a diagnostic tool for the CP sensitivity of each TQD molecule. To obtain the re\-flec\-ti\-vities and the Kerr angle we calculate the
conduc\-ti\-vi\-ty ($\bm{\sigma}$) and dielectric ($\bm{\varepsilon}$) tensors \cite{Kerr,Sup}, which provide the complex refractive index. The two tensors are related
via $\bm{\varepsilon}(\omega)=\bm{1}+i\bm{\sigma}(\omega)/\omega\varepsilon_0$, where $\varepsilon_0$ defines the permittivity of vacuum. The $C_3$ symmetry imposes
$\sigma_{xx}(\omega)=\sigma_{yy}(\omega)\equiv\sigma(\omega)$ and $\sigma_{xy}(\omega)=-\sigma_{yx}(\omega)\equiv\sigma_H(\omega)$. For the calculations, we consider zero
tem\-pe\-ra\-ture and assume that only the state $\left|0\right>$ is initially occupied, allowing transitions to $\left|\pm\right>$ with Bohr frequencies $\omega_{\pm,0}$. 
We find that for high frequencies, $\omega\gg\omega_{\pm,0}$, the dielectric functions for each CP ($\varepsilon_{\pm}$) take the form $\varepsilon_{\pm}(\omega)\simeq
1+[\pm{\rm Re}\ph \sigma_{H}(\omega)-{\rm Im}\ph \sigma(\omega)]/\omega\varepsilon_0$.

\textit{CP-birefringence.} In Fig.~\ref{Fig:KerrPumping}(a) we present results for the reflectivities $R_{\pm}$, with applied flux corresponding to $\nu=1/6$ yielding
$\hbar\omega_{+,0}=34{\rm \mu eV}$ and $\hbar\omega_{-,0}=52{\rm \mu eV}$. The reflectivities exhibit small dips at the absorption lines $\omega_{\pm,0}$.
In this frequency region, the TQD molecule becomes highly reflective, with a small diffe\-ren\-ce $\delta R\equiv R_--R_+$. For photon frequencies above the threshold
frequency, $\Omega\simeq (\omega_{-,0}^3+\omega_{+,0}^3)/(\omega_{-,0}^2-\omega_{+,0}^2)$ (see \cite{Sup}), $\delta R$ changes dramatically and strong birefringence sets in.
The sharp dip at $\Omega$ occurs due to complete transparency of the TQD molecule to RHCP, i.e. $R_+=0$, while at the same time the LHCP is completely reflected, i.e.
$R_-\simeq1$. The reason for this sharp dip is that the dielectric function becomes $\varepsilon_+(\Omega)=1$, since at this frequency ${\rm Re}\ph \sigma_{H}(\omega)={\rm
Im}\ph \sigma(\omega)$. Obviously, this feature requires a finite Hall conductivity and thus emerges only in the presence of a magnetic field.

\textit{Tunable Kerr effect.} The TQD can additionally produce a Kerr angle $\theta_K$ of measurable size, rea\-ching a few ${\rm mrads}$ even for \textit{weak} magnetic
fields. In Fig.~\ref{Fig:KerrPumping}(b) we show results for $\nu=1/20$ with $\hbar\omega_{+,0}=44{\rm \mu eV}$ and $\hbar\omega_{-,0}=50{\rm \mu eV}$. The Kerr angle is in
principle observable for $\omega>\omega_{+,0}$ and is characterized by a sharp peak-like feature for $\omega\simeq\omega_{-,0}$. For $\omega \gg \omega_{-,0}$ the Kerr angle
increases monotonically and acquires quite high va\-lues. Remarkably, in the region where both $\delta R$ and $\theta_K$ become significant and experimentally acessible, the
TQD molecule can be employed as a sensitive CP-filter, rotator and detector.

\textit{CP-sensitive lasing.}  The strategy for lasing action based on TQD molecules is to operate close to an energy de\-ge\-ne\-racy and to create a population inversion
between two nearly dege\-ne\-ra\-te states. This allows controlling the han\-ded\-ness of the emission CP and inverting it by a slight modification of the magnetic field. For
instance, a la\-sing transition between the states $\left|+\right>$ and $\left|0\right>$ at $\nu = 0.40$ couples to the photons with RHCP, as indicated in Fig.~\ref{Fig:fig2}.
For an interdot distance of $200 {\rm nm}$ an increase of the magnetic field of about $50 \, {\rm mT}$ brings the system to $\nu=0.64$, where the photons with the same
frequency ($f \simeq 2\, {\rm GHz}$) but LHCP couple to the lasing states.

In order to create the population inversion $\tau_0$ \cite{PI,Sup}, we propose a pumping scheme based on the intrinsic three le\-vels of the TQD. For definiteness we
consider the level arrangement of Fig.~\ref{Fig:Dipole}(a). An unpolarized broadband radiation pumps from $\left|0\right>$ to $\left|-\right>$, while the la\-sing transition
involves the states $\left|+\right>$ and $\left|0\right>$ (Fig.~\ref{Fig:KerrPumping}(c)). For simplicity we assume a classical pumping field that induces up-
and down-transitions with equal rates $\Gamma_{\rm p}$. Po\-pu\-lation inversion occurs when the rela\-xa\-tion rate $\Gamma_1$ from the state $\left|-\right>$ to
$\left|+\right>$ is larger than the rate $\Gamma_2$ from $\left|+\right>$ to $\left|0\right>$, in combination with the condition $\Gamma_{\rm p}>
\Gamma_1\Gamma_2/(\Gamma_1-\Gamma_2)$. Since in the particular case all the molecular transitions occur with equal matrix e\-le\-ments (Eq.~\eqref{eq:coupling}) a difference
between $\Gamma_{1,2}$ can be achieved via an asymmetric coupling to the environment, which arises if $\omega_{-,+}\neq\omega_{+,0}$. In a rea\-li\-stic si\-tua\-tion, the
broadband pum\-ping field may also effect the transition between $\left|-\right>$ and $\left|+\right>$ due to their closeness in frequencies, while the state $\left|-\right>$
also has a finite re\-la\-xa\-tion rate $\Gamma_3$ to the state $\left|0\right>$. Both sources reducing the population inversion can be avoided by an appropriate choice of the
pumping field frequency. Note also that the aforementioned conditions are not overly restrictive, as lasing can occur even without population inversion \cite{Michael}.

Another aspect which is important for the CP-sensitive applications is the strength $g$, of the coupling to the ca\-vi\-ty. Strong coupling has been demonstrated for both
gate-defined and self-assembled QDs. Speci\-fi\-cal\-ly, for gate-defined QDs coupled to a microwave transmission line cavity, the coupling strength $g$ can reach tens of MHz
for an interdot distance of about a hundred nanometers \cite{QDCoup1,QDCoup2}. In addition, strong coupling of the order of 10 GHz was recently reported for a single
self-assembled QD coupled to a fiber Fabry-Perot cavity \cite{SAQD}. The expe\-ri\-men\-tally feasible strong coupling to the cavity allows satisfying the additional lasing
condition $g\geq \sqrt{\kappa(\Gamma_{\varphi}^2+\Delta^2)/(2\tau_0\Gamma_{\varphi})}$, which expresses that the coupling has to be strong to overcome the total de\-phasing of
the QDs with: rate $\Gamma_{\varphi}$, the damping of the cavity with rate $\kappa$, and the frequency detuning $\Delta$ between the dot states and the cavity \cite{SQL}.

The ope\-ra\-tional frequency range of the proposed device is determined by the energy splitting of the TQD molecule, which may reach few tens of GHz. In addition, the CP of
the emission spectrum depends on the energy-level arrangement, which for the typical realization can be manipulated and switched by changing a \textit{weak} magnetic field by
a few tens of ${\rm mT}$. Further miniaturization of the interdot distance, down to the order of twenty nanometers, opens perspectives for applications and la\-sing operation
in the lower far infrared regime. However, the degree of CP-tunability is reduced in this case, since the small TQD area requires a magnetic field above $5{\rm
T}$ to change the flux quanta by ten percent.

The realistic design of our proposal requires esti\-ma\-ting the effects of possible imperfections, e.g. stray in-plane electric fields and misaligned QDs, that both lead to
$C_3$ symmetry breaking. For weak electric and displacement fields, we proceed perturbatively and find \cite{Sup} that: \bt{i.} $t$ is modified and \bt{ii.} each molecular
transition between the perturbed eigenstates can occur via emitting/absorbing photons of \textit{both} CPs, with emission probabilities $\tilde{{\rm p}}_{\pm}$. Notably, the
degree of CP sen\-si\-ti\-vi\-ty: $(\tilde{{\rm p}}_+-\tilde{{\rm p}}_-)/(\tilde{{\rm p}}_++\tilde{{\rm p}}_-)$, remains unchanged at this level of approximation. However, we
obtain mixed CPs ($\widetilde{{\cal E}}_{\pm}$), which nevertheless still exhibit a frequency region of strong birefringence. For an interdot distance $200$nm, $\omega=\Omega$
and $\nu=1/6$, an 8.7nm displacement of a single QD or an electric field of strength $0.45$kV/m, yield a $9\%$ CP admixture.

Our quantum metamaterial may also support inter\-mo\-le\-cu\-lar interference effects, which are observable when the molecules get closer than appro\-xi\-mately 400nm. For
larger separations the molecules are sufficiently decoupled and the above analysis is directly applicable. Our results can be also generalized to other equi\-la\-te\-ral
polygon geometries. In this case additional mo\-le\-cu\-lar transitions are accessible, while the emission spectrum can be tailored by properly selecting the polygon type.

In summary, we proposed a new highly tunable CP-sensitive metamaterial based on arrays of artificial molecules. Each molecule consists of three triangularly arranged
quantum-dots. The particular platform is in principle accessible in the lab and can be readily employed for CP-sensitive applications in the microwave regime, taking advantage
of the high-tunability provided by the quantum-dot technology. The system can show strong signatures of CP-birefringence, which become e\-vi\-dent in the reflectivity and the
Kerr rotation, rendering the setup a CP-filter and detection device. Finally, we showed that lasing action accompanied by CP-conversion is realizable in this frequency domain
and can be finely tuned by weak magnetic fields. 

We would like to thank J. Cole, Y. Utsumi, T. \v{C}ade\v{z}, A. Ram\v{s}ak, T. Rejec and A. Kregar for numerous fruitful discussions. P.Q.J. was financially supported by the
National Natural Science Foundation of China (Grant No. 11304196) and the Science and Technology Program of Shanghai Maritime University. P.K and G. S. acknowledge funding
from the EU project NanoCTM (No. 234970).

\begin{widetext}
\newpage

\appendix

\section*{{\Large Supplementary Material}}

\section{A. Estimation of the interdot tunneling strength}\label{Sec:ES}

To estimate the  tunneling strength we use a microscopic model of a lateral double-dot system occupied by one electron. It allows for simple analytical results. 
The quantum dots are located at $\mathbf R_{\rm L/R} = (\pm \sqrt{3}\,a/2, 0)$. A perpendicular magnetic field,  $\bm{{\cal B}}={\cal B}_z\bm{\hat{z}} $, is applied,
for which we adopt a symmetric gauge $\mathbf A(\mathbf r)=\displaystyle\frac{{\cal B}_z}{2}(-y,x,0)$. The total Hamiltonian can be written as a sum of an orbital part,
$H_{\rm orb}=\frac{1}{2m} \Bigl[\mathbf P + e \mathbf A(\mathbf r) \Bigr]^2 + V_{\rm ovl}(\mathbf r)$, and a Zeeman term, $H_{\rm Z}$. Here we adopt an overall confinement by
\cite{Burkard99},
\begin{eqnarray}\label{Eq:Confinement}
V_{\rm ovl}(\mathbf r) = \frac{m\,\omega_0^2}{2}
            \Bigl[\frac{1}{3 a^2} \Bigl(x^2-\frac{3 a^2}{4}\Bigr)^2+y^2 \Bigr],
\end{eqnarray}
which reduces to a parabolic potential well $V_{\rm L/R}(\mathbf r) = m\,\omega_0^2 \Bigl[(x\pm \sqrt{3}a/2)^2 +y^2 \Bigr]/2 $ around each dot center $\mathbf R_{\rm L/R}$. 
We employ a tight-binding model with Hamiltonian of each isolated quantum dot given by
\begin{eqnarray}
 h_{\rm L/R} = \frac{1}{2m} \Bigl(\mathbf P + e \mathbf A \Bigr)^2 + V_{\rm L/R},
\end{eqnarray}
whose eigenstates are known as Fock-Darwin states \cite{FDStates}. We assume strong confinement, leading to large energy splitting of several hundreds of $\rm \mu eV$.
Hence we only focus on the degenerate ground states of the two isolated quantum dots,
\begin{eqnarray}\label{Eq:SngDotOrb}
 \langle \mathbf r| {\rm L/R}\rangle \equiv \psi_{\rm L/R}(\mathbf r) = \frac{1}{\sqrt{\pi} l}
        \exp \Bigl[-\frac{(x\pm \sqrt{3}a/2)^2 + y^2}{2\, l^2} \pm i\frac{\sqrt{3}a}{4\, l_{\rm B}^2}y   \Bigr],
\end{eqnarray}
with ground-state energy $\hbar\,\omega=\hbar\sqrt{\omega_0^2+(e{\cal B}_z/2m)^2}$. Here $l = \sqrt{\hbar/(m\omega)}$ represents the effective radius of the
electronic orbitals and $l_{\rm B} = \sqrt{\hbar /(e{\cal B}_z)}$ the magnetic length. The two orbitals have an overlap $S=\exp\left[-d^2(1+2\xi_{\rm
d}^2\,\nu^2)/(4\sqrt{1+\xi_{\rm d}^2\,\nu^2}) \right]$, where $d=\sqrt{3}\,a/a_{\rm B}^{}$ denotes the interdot distance scaled by the Bohr radius $a_{\rm B}^{} =
\sqrt{\hbar/(m\omega_0)}$, and the coefficient $\xi_{\rm d} = 4\pi/(\sqrt{3}\,d^2)$ accounts for the geometry of the TQD. For the comparison to the triple quantum dot 
molecule we express the dependence on the magnetic field by the normalized flux $\nu=\Phi/\Phi_0 = 3\sqrt{3}\,e{\cal B}_z a^2/(4 h)$.

We then orthogonalize the dot states via  $|\Psi_{\rm L/R}\rangle = N_{\rm m}\Bigl(|{\rm L/R}\rangle - \alpha |{\rm R/L}\rangle\Bigr)$, where $\alpha = (1-\sqrt{1-S^2})/S$ 
represents the correction due to the neighboring dot, and the normalization factor is $N_{\rm m} = (1-2\alpha S + \alpha^2)^{-1/2}$. For weak overlap $\alpha \approx S/2$. In
the space spanned by $\{|\Psi_{\rm L}\rangle, |\Psi_{\rm R}\rangle \}$, the Hamiltonian is given by $H_{\rm orb} = \epsilon\,I+t\,\sigma_x $,
with
\begin{eqnarray}
\epsilon&=& \hbar\,\omega_0 \left[ \sqrt{1+\xi_{\rm d}^2\,\nu^2}
+ \frac{3}{8(1+\xi_{\rm d}^2\,\nu^2)d^2} - \frac{3S^2}{8(1-S^2)} \left(1+\frac{d^2}{4}\right)   \right],
            \no \\[3mm]
 t &=& \frac{3\,\hbar\,\omega_0}{8} \frac{S}{1-S^2}\left(\frac{1}{\sqrt{1+\xi_{\rm d}^2\,\nu^2}} + \frac{d^2}{4}  \right),\label{eq:Bdep}
\end{eqnarray}

\noindent and $\sigma_x = |\Psi_{\rm L}\rangle\langle \Psi_{\rm R}|+|\Psi_{\rm R}\rangle\langle \Psi_{\rm L}|$ being the Pauli matrix as well as $I$ the identity matrix.
When the overlap $S$ is small and the interdot distance is of the same order of the Bohr radius $a_{\rm B}$, the tunneling strength approximates $S\hbar\omega_0 $.

\section{B. Conductivity tensor}\label{Sec:Conductivity}

\noi For the calculation of the conductivity tensor we consider the  current-current correlation function and take into account only the lowest-order bubble diagram:
\bea
\sigma_{ss'}(\omega)=\sum_{\lambda,\lambda'}^{0,\pm}
\frac{i\left<\lambda\right|\widehat{J}_{s}\left|\lambda'\right>\left<\lambda'\right|\widehat{J}_{s'}\left|\lambda\right>\left[f(E_{\lambda})-f(E_{\lambda'})\right]}
{v\hbar\omega\left(\omega_{\lambda,\lambda'}+\omega+i\eta\right)}\,,
\label{eq:conductivity}
\eea

\noi where $f(E)$ denotes the Fermi-Dirac distribution, $s,s'=x,y$ and $\eta\rightarrow0^+$. Here we assumed thermal equilibrium. We introduced the effec\-tive vo\-lu\-me of
the TQD molecule, $v=\pi a^2w$, with $w\sim10{\rm nm}$ corresponding to the diameter of each quantum-dot \cite{dz}. In Eq.~\eqref{eq:conductivity},
$\widehat{\bm{J}}$ defines the current operator, which is related to the polarization operator of the TQD molecule through the relation
$\widehat{\bm{J}}=\partial\widehat{\bm{P}}/\partial t$. The current operator is directly obtained from the expression
$\widehat{\bm{J}}=i[\widehat{{\cal H}}_{\rm TQD},\widehat{\bm{P}}]/\hbar$ yielding the matrix elements
$\left<\lambda\right|\widehat{\bm{J}}\left|\lambda'\right>=i\omega_{\lambda,\lambda'}\left<\lambda\right|\widehat{\bm{P}}\left|\lambda'\right>$ with $\lambda,\lambda'=0,\pm$.
The eigenenergies differences are compactly written as
\bea
\hbar\omega_{\lambda,\lambda'}=E_{\lambda}-E_{\lambda'}=-4t\sin\left[(\lambda-\lambda')\pi/3\right]\sin\left[\phi-(\lambda+\lambda')\pi/3\right]\,.
\eea

\noi The conductivity tensor satisfies $\sigma_{xx}(\omega)=\sigma_{yy}(\omega)\equiv\sigma(\omega)$ and $\sigma_{xy}(\omega)=-\sigma_{yx}(\omega)\equiv\sigma_{H}(\omega)$, 
the latter denoting the Hall conductivity. The general expressions for the conductivities, for $\omega>0$ and $\eta\rightarrow0^+$ (for the actual calculation of the
conductivity tensor we used $\hbar\eta=0.3{\rm \mu eV}$), become
\bea
\sigma(\omega)&=&-\frac{e^2}{wh}\sum_{(\lambda,\lambda')}^{{\color{black}\bm{\circlearrowleft}}}
\frac{f\left(E_{\lambda}\right)-f\left(E_{\lambda'}\right)}{\omega}
\frac{\omega_{\lambda,\lambda'}^3}{\omega+\left|\omega_{\lambda,\lambda'}\right|}
\frac{\eta+i\left(\omega-\left|\omega_{\lambda,\lambda'}\right|\right)}
{\left(\omega-\left|\omega_{\lambda,\lambda'}\right|\right)^2+\eta^2}\,,\\\no\\
\sigma_H(\omega)&=&-\frac{e^2}{wh}\sum_{(\lambda,\lambda')}^{{\color{black}\bm{\circlearrowleft}}}
\frac{f\left(E_{\lambda}\right)-f\left(E_{\lambda'}\right)}{\omega_{\lambda,\lambda'}}
\frac{\omega_{\lambda,\lambda'}^3}{\omega+\left|\omega_{\lambda,\lambda'}\right|}
\frac{\omega-\left|\omega_{\lambda,\lambda'}\right|-i\eta}
{\left(\omega-\left|\omega_{\lambda,\lambda'}\right|\right)^2+\eta^2}\,,
\eea

\noi where ${\color{black}\bm{\circlearrowleft}}\equiv\{(0,+),(+,-),(-,0)\}$ denotes the pairs of $(\lambda,\lambda')$ that have to be taken into account. If we assume
zero-temperature and therefore consider that only the $\left|0\right>$ is occupied, we obtain the Bohr frequencies $\hbar\omega_{\lambda,0}\equiv
E_{\lambda}-E_{0}=2\sqrt{3}t\sin\left(\pi/3-\lambda \phi\right)=2\sqrt{3}t\sin\left[(1-2\lambda\nu)\pi/3\right]\geq0$ and the conductivities
\bea
\sigma(\omega)=\frac{e^2}{wh}\sum_{\lambda=\pm}\frac{\omega_{\lambda,0}^3}{\omega(\omega_{\lambda,0}+\omega)}
\frac{\eta-i\left(\omega_{\lambda,0}-\omega\right)}{\left(\omega_{\lambda,0}-\omega\right)^2+\eta^2}\quad{\rm and}\quad
\sigma_H(\omega)=\frac{e^2}{wh}\sum_{\lambda=\pm}\frac{\lambda\omega_{\lambda,0}^2}{\omega_{\lambda,0}+\omega}
\frac{(\omega_{\lambda,0}-\omega)+i\eta}{\left(\omega_{\lambda,0}-\omega\right)^2+\eta^2}\,.
\eea

\noi More compactly, we write $\sigma(\omega)=\sum_{\lambda}\sigma^{\lambda}(\omega)$ and $\sigma_H(\omega)=\sum_{\lambda}\sigma_H^{\lambda}(\omega)$, with $\lambda=\pm$.
Note that $\sigma_H(\omega)=0$ for $\nu=3k/2$ with $k\in\mathbb{Z}$, since $\omega_{+,0}=\omega_{-,0}$, implying that ${\cal T}$ is restored. For frequencies
$\omega\gg\omega_{\pm,0}$, the conductivities become
\bea
\sigma(\omega)\simeq\frac{e^2}{wh}\frac{i\left(\omega_{-,0}^3+\omega_{+,0}^3\right)}{\omega^3}\quad{\rm and}\quad
\sigma_H(\omega)\simeq\frac{e^2}{wh}\frac{\omega_{-,0}^2-\omega_{+,0}^2}{\omega^2}\,.
\eea

\section{C. Dielectric tensor and CP birefringence}\label{Sec:Dielectric}

\noi Starting from the conductivity tensor we calculate the dielectric tensor $\bm{\varepsilon}$ through the defining relation
\bea
\bm{\varepsilon}(\omega)=\bm{1}-\frac{\bm{\sigma}(\omega)}{i\omega\varepsilon_0}=
\bm{1}-\frac{{\rm Im}\ph \bm{\sigma}(\omega)}{\omega\varepsilon_0}+i\frac{{\rm Re}\ph \bm{\sigma}(\omega)}{\omega\varepsilon_0}\,.
\eea

\noi Here we have introduced the permittivity of vacuum $\varepsilon_0$ and considered electromagnetic waves of the  form $\bm{{\cal E}}(z,t)=\bm{{\cal E}} 
e^{i(qz-\omega t)}$. By introducing
\bea
\varepsilon(\omega)=1-\frac{{\rm Im}\ph \sigma(\omega)}{\omega\varepsilon_0}+i\frac{{\rm Re}\ph \sigma(\omega)}{\omega\varepsilon_0}\qquad{\rm and}\qquad
\varepsilon_H(\omega)=\frac{\sigma_{H}(\omega)}{\omega\varepsilon_0}\,,
\eea

\noi we can diagonalize the dielectric tensor and obtain the dispersions
\bea
\omega=\frac{cq}{\sqrt{\varepsilon_{\pm}(\omega)}}\,,
\eea

\noi with $\varepsilon_{\pm}(\omega)=\varepsilon(\omega)\pm\varepsilon_H(\omega)$ and complex refractive index, $N_{\pm}(\omega)=\sqrt{\varepsilon_{\pm}(\omega)}$,
corresponding to the circular polarizations ${\cal E}_{\pm}$. Consequently the TQD exhibits optical birefringence and we expect a finite Kerr angle for the reflected beam of
an incident linearly polarized beam onto the TQD. For $\omega\gg\omega_{\pm,0}$ the dielectric tensor takes the approximate form
\bea
\varepsilon_{\pm}(\omega)\simeq 1+\frac{1}{\omega\varepsilon_0}\left[\pm{\rm Re}\ph \sigma_H(\omega)-{\rm Im}\ph \sigma(\omega)\right]=
1+\frac{e^2}{\omega\varepsilon_0 hw}\left[\pm\frac{\omega_{-,0}^2-\omega_{+,0}^2}{\omega^2}-\frac{\omega_{-,0}^3+\omega_{+,0}^3}{\omega^3}\right]\,.\label{eq:condApprox}
\eea

\noi Note that $\varepsilon_{+}=1$ for frequencies near $\Omega=\left(\omega_{-,0}^3+\omega_{+,0}^3\right)/\left(\omega_{-,0}^2-\omega_{+,0}^2\right)$, leading
to complete transparency for the right-handed CP, while the other is almost perfectly reflected.

\section{D. Reflectivity and Kerr angle}\label{Sec:ReflectivityKerr}

\noi We consider  an electromagnetic wave of the form
\bea
\bm{{\cal E}}_i(z,t)=+{\cal E}_{i,+}\hat{\bm{e}}_+ e^{i(qz-\omega t)}+{\cal E}_{i,-}\hat{\bm{e}}_- e^{i(qz-\omega t)}\,
\eea

\noi incident on the TQD with $cq=\omega$. Here $\hat{\bm{e}}_{\pm}$ correspond to the unit vectors for the respective CP field. Due to the interface to the TQD there is a
reflected and transmitted beam of the  form
\bea
\bm{{\cal E}}_r(z,t)&=&-{\cal E}_{r,+}\hat{\bm{e}}_+ e^{-i(qz+\omega t)}-{\cal E}_{r,-}\hat{\bm{e}}_- e^{-i(qz+\omega t)}\,,\\
\bm{{\cal E}}_t(z,t)&=&+{\cal E}_{t,+}\hat{\bm{e}}_+ e^{i(q_+z-\omega t)}+{\cal E}_{t,-}\hat{\bm{e}}_- e^{i(q_-z-\omega t)}\,,
\eea

\noi with $cq_{\pm}=\omega\sqrt{\varepsilon_{\pm}(\omega)}=\omega N_{\pm}(\omega)$. By introducing the reflection and transmission coefficients ${\cal E}_{r\pm}=r_{\pm}{\cal
E}_{i\pm}$ and ${\cal E}_{t\pm}=t_{\pm}{\cal E}_{i\pm}$, respectively, and considering the continuity of the fields at the interface $z=0$ we obtain
\bea
1-r_{\pm}=t_{\pm}\qquad{\rm and}\qquad
1+r_{\pm}&=&t_{\pm}N_{\pm}(\omega)\,,
\eea

\noi which yields
\bea
r_{\pm}(\omega)=\frac{N_{\pm}(\omega)-1}{N_{\pm}(\omega)+1}\qquad{\rm and}\qquad
t_{\pm}(\omega)=\frac{2}{N_{\pm}(\omega)+1}\,.
\eea

\noi The reflectivity reads $R_{\pm}(\omega)=|r_{\pm}(\omega)|^2$. The Kerr angle \cite{Kerr} is given by the expression $\theta_K=(\theta_+-\theta_-)/2$,
with the related angles defined as
\bea
\theta_l=\tan^{-1}\left[\frac{{\rm Im}\ph r_l(\omega)}{{\rm Re}\ph r_l(\omega)}\right]\,,\quad{\rm with}\phd l=\pm.
\eea

\section{E. Lasing and population inversion}\label{Sec:PopInversion}

\noindent The dynamics of the system is described by the master equation for the density matrix $\rho$ \cite{PI},
\begin{eqnarray}
 \dot{\rho} = -\frac{i}{\hbar}[H_{\rm sys},\rho] + \sum_i \mathcal{L}_i \rho,
\end{eqnarray}
where $H_{\rm sys}$ denotes the Hamiltonian for the coupled dot-cavity system, and $\mathcal L_i$ the Liouville superoperators for dissipation. Here we assume the system is
weakly coupled to the environment and adopt the Born-Markovian approximation where the dissipative dynamics are described in the Lindblad form,
\begin{eqnarray}
\mathcal L_i \rho = \frac{\Gamma_i}{2} \left(2 L_i \rho L_i^\dag - L^\dag_i L_i \rho -\rho L^\dag_i L_i \right).
\end{eqnarray}

For definiteness, we consider the situation in Fig.~4(c) where the cavity is resonant to the transition between the states $|+\rangle$ and $|0\rangle$, while the pumping field
connects the states $|-\rangle$ and $|0\rangle$. In this case, the relaxation and excitation induced by the incoherent pumping are described by $L_{\uparrow} =
\left|-\right>\left<0\right|$ and $L_{\downarrow}=\left|0\right>\left<-\right|$ with the same rate $\Gamma_{\rm p}$, and the relaxation processes
$\left|-\right>\rightarrow\left|+\right>$ and $\left|+\right>\rightarrow\left|0\right>$ by $L_1 = \left|+\right>\left<-\right|$ and $L_2 = \left|0\right>\left<+\right|$
with rates $\Gamma_{1,2}$, respectively. The population inversion, $\tau_0$, can then be obtained from the steady state solution of the master equation with vanishing coupling
to the cavity, which is given by
\bea
\label{Eq:PI}
\tau_0 =\frac{\gamma_1 \gamma_{\rm p}-\gamma_1-\gamma_{\rm p}}{\gamma_1 \gamma_{\rm p}+\gamma_1+\gamma_{\rm p}}\,,
\eea

\noindent with rates $\gamma_{1, \rm{p}} = \Gamma_{1, \rm{p}} /\Gamma_2$. From Eq.~\eqref{Eq:PI}, a positive population inversion is achieved when the relaxation from the
state $|-\rangle$ is stronger than that from $\left|+\right>$, namely, $\gamma_1 >1$, in combination with a strong pumping field satisfying 
$\gamma_{\rm p}>\gamma_1/(\gamma_1-1)$.

\section{F. Effects of broken $\bm{C_3}$ symmetry}\label{Sec:Asymmetry}

In this section we discuss the consequences of a weakly broken $C_{3}$ symmetry, either due to the presence of: \bt{i.} a stray in-plane electric field, $\bm{E}$ or
\bt{ii.} asymmetrically positioned quantum dots comprising the TQD molecule. The latter $C_3$ symmetry breaking sources modify the on-site dot energies:
$\epsilon\rightarrow\epsilon_i$, the hopping matrix elements $t\rightarrow$ $t_{ii'}$ (with $i,i'=1,2,3$) and the polarization operator
$\widehat{\bm{P}}\rightarrow\widehat{\bm{P}}+\delta\widehat{\bm{P}}$. In particular, $\bm{E}$ only modifies the on-site energies due to the inhomogeneity of the
electrostatic landscape. Here we consider that the quantum dot located at position $\bm{R}_1$ is displaced by $\delta\bm{R}_1=a(u_x,u_y)$, with $\bm{u}$ the dimensionless
displacement vector. Thus its new position is $\bm{R}_1+\delta\bm{R}_1=a(1+u_x,u_y)$. The field $\bm{u}$ contributes to the modification of the three aforementioned
quantities. For the present discussion we consider that the differences $\epsilon-\epsilon_i$ and $t-t_{ii'}$ are small, allowing us to proceed perturbatively. In addition, we
assume the case of a non-vanishing ${\cal B}_z$ field leading to an orbital level splitting larger than the above-mentioned energy differences. Under these conditions we write
\bea
\epsilon_1=\epsilon+\delta\epsilon_x\,,\phd
\epsilon_2=\epsilon-\frac{\delta\epsilon_x-\sqrt{3}\delta\epsilon_y}{2}\,,\phd
\epsilon_3=\epsilon-\frac{\delta\epsilon_x+\sqrt{3}\delta\epsilon_y}{2}\phd\Rightarrow\phd
\delta\epsilon_x=\frac{\epsilon+\epsilon_1-\epsilon_2-\epsilon_3}{2}\phd{\rm and}\phd \delta\epsilon_y=\frac{\epsilon_2-\epsilon_3}{\sqrt{3}}\,,\quad\\
t_{12}=\tilde{t}-\frac{\delta t_x}{2}+\frac{\delta t_y}{2}\sqrt{3}\,,\phd
t_{13}=\tilde{t}-\frac{\delta t_x}{2}-\frac{\delta t_y}{2}\sqrt{3}\,,\phd 
t_{23}=\tilde{t}+\delta t_x\Rightarrow
\delta t_x=\frac{2t_{23}-t_{12}-t_{13}}{3}\phd{\rm and}\phd\delta t_y=\frac{t_{12}-t_{13}}{\sqrt{3}}\,,\quad
\eea

\noi while the displacement of the dot introduces the following modification on the polarization operator $\delta\widehat{\bm{P}}=-e\left|1\right>\bm{R}_1\left<1\right|$. Note
that \textit{also} the $C_3$ symmetric tunneling amplitude $t$ becomes generally modified $t\rightarrow \tilde{t}=(t_{12}+t_{13}+t_{23})/3$. Using the expressions for the
on-site energy $\epsilon$ and the coherent tunneling $t$ (see Eq.~\ref{eq:Bdep}), one finds that $\delta\epsilon_x=eaE_x+\zeta au_x$, $\delta\epsilon_y=eaE_y-\zeta au_y$,
$\delta t_x=\kappa au_x$ and $\delta t_y=\kappa au_y$. The value $\tilde{t}$, as also the coupling constants $\zeta$ and $\kappa$ can be calculated from Eq.~\eqref{eq:Bdep} by
taking into account that the interdot distances become $d_{12}\simeq d+ d(u_x-u_y/\sqrt{3})/2$, $d_{13}\simeq d+d(u_x+u_y/\sqrt{3})/2$ and $d_{23}=d$. Assuming an interdot
distance $200$nm we find $\xi_{\rm d}^2\simeq1/5$. In addition, for the value of normalized flux considered in the main text, $\nu=1/6$, we see that $1+\xi_{\rm
d}^2\nu^2\simeq1$. Since we have a weak overlap $S\ll 1$, we find $\zeta\simeq0$, $\kappa a=(2t/3)(d/2)^4/[1+(d/2)^2]$ and $\tilde{t}=t-t(2u_x/3)(d/2)^4/[1+(d/2)^2]$.

\subsection{Eigenstate mixing}

In the presence of $\bm{E}$ and $\bm{u}$, the $C_3$ symmetric Hamiltonian presented in the main text, $\widehat{{\cal H}}_{\rm TQD}(t)$ becomes $\widehat{{\cal H}}_{\rm
TQD}(\tilde{t})$, and also acquires the gauge-invariant $C_3$ symmetry-breaking part
\bea
\widehat{{\cal V}}
=\frac{\delta\epsilon_x}{2}\left(\begin{array}{ccc}2&0&0\\0&-1&0\\0&0&-1\end{array}\right)
+\frac{\delta\epsilon_y}{2}\sqrt{3}\left(\begin{array}{ccc}0&0&0\\0&1&0\\0&0&-1\end{array}\right)
+\frac{\delta t_x}{2}\left(\begin{array}{ccc}0&e^{i\phi}&e^{-i\phi}\\e^{-i\phi}&0&-2e^{i\phi}\\e^{i\phi}&-2e^{-i\phi}&0\end{array}\right)
-\frac{\delta t_y}{2}\sqrt{3}\left(\begin{array}{ccc}0&e^{i\phi}&-e^{-i\phi}\\e^{-i\phi}&0&0\\-e^{i\phi}&0&0\end{array}\right)\,.\quad
\eea

\noi Here we retained only the linear order in the $\bm{E}$ and $\bm{u}$ fields. Note that $\widehat{{\cal V}}$ is expressed in the dot basis. By applying non-degenerate
perturbation theory at first order we obtain the corrected eigenenergies and eigenstates
\bea
\widetilde{E}_{\lambda}\simeq E_{\lambda}+\left<\lambda\right|\widehat{{\cal V}}\left|\lambda\right>\simeq E_{\lambda}(t\rightarrow \tilde{t})\qquad{\rm and}\qquad
\big|\tilde{\lambda}\big>\simeq\left|\lambda\right>+\sum_{\lambda'\neq\lambda}
\frac{\left<\lambda'\right|\widehat{{\cal V}}\left|\lambda\right>}{\hbar\omega_{\lambda,\lambda'}}\left|\lambda'\right>\equiv
\left|\lambda\right>+\sum_{\lambda'\neq\lambda}{\cal M}_{\lambda,\lambda'}\left|\lambda'\right>\,,
\eea

\noi where ${\cal M}_{\lambda,\lambda'}$ define the mixing matrix elements. The latter satisfy ${\cal M}_{\lambda',\lambda}=-{\cal M}_{\lambda,\lambda'}^*$. The matrix
representation of the symmetry breaking Hamiltonian, in the chiral basis, reads
\bea
\widehat{{\cal V}}=
\frac{1}{\sqrt{2}}\left(\begin{array}{ccc}0&\delta\epsilon_-&\delta\epsilon_+\\\delta\epsilon_+&0&\delta\epsilon_-\\\delta\epsilon_-&\delta\epsilon_+&0\end{array}\right)
-\frac{\kappa a}{\sqrt{2}}\left(\begin{array}{ccc}0&2\cos\left(2\pi/3+\phi\right)u_+&2\cos\left(2\pi/3-\phi\right)u_-
\\2\cos\left(2\pi/3+\phi\right)u_-&0&2\cos\phi\ph u_+\\2\cos\left(2\pi/3-\phi\right)u_+&2\cos\phi\ph u_-&0\end{array}\right)\,,
\eea

\noi where we introduced the right/left handed fields $\delta\epsilon_{\pm}=(\delta\epsilon_x\pm i\delta\epsilon_y)/\sqrt{2}=ea E_{\pm}+\zeta au_{\mp}$ and $u_{\pm}=(u_x\pm i
u_y)/\sqrt{2}$. From the above we obtain the mixing matrix elements in the chiral basis:
\bea
{\cal M}_{\pm,0}=\frac{{\cal V}_{0,\pm}}{\hbar\omega_{\pm,0}}=\frac{a}{\sqrt{2}}
\frac{eE_{\mp}+\zeta u_{\pm}-2\cos(2\pi/3\pm\phi)\kappa u_{\pm}}{2\sqrt{3}t\sin(2\pi/3\pm\phi)}
\quad{\rm and}\quad
{\cal M}_{-,+}=\frac{{\cal V}_{+,-}}{\hbar\omega_{-,+}}=\frac{a}{\sqrt{2}}\frac{eE_-+\zeta u_+-2\cos\phi\kappa u_+}{2\sqrt{3}t\sin\phi}\,.\phd
\eea

\noi Conclusively, we observe that weak breaking of $C_3$ symmetry modifies the emission spectrum due to $t\rightarrow \tilde{t}$. More importantly, the symmetry breaking
perturbations mix the chiral basis eigenstates.

\subsection{Polarization operator}

The expression of the polarization operator becomes modified due to the displacement of the quantum-dot located at $\bm{R}_1$. At first order in the $C_3$ breaking
perturbations, we obtain the following modified right handed polarization operator matrix
\bea
\widehat{P}_++\delta\widehat{P}_+&\simeq&
-\frac{ea}{\sqrt{2}}\left(\begin{array}{ccc}
0&1&0\\
0&0&1\\
1&0&0
\end{array}\right)
-\frac{eau_+}{3}\left(\begin{array}{ccc}1&1&1\\1&1&1\\1&1&1\end{array}\right)
-\frac{ea}{\sqrt{2}}\left(\begin{array}{ccc}
{\cal M}_{0,+}+{\cal M}_{0,-}^*&0&{\cal M}_{0,+}^*+{\cal M}_{-,+}\\
{\cal M}_{0,-}+{\cal M}_{+,-}^*&{\cal M}_{+,-}+{\cal M}_{+,0}^*&0\\
0&{\cal M}_{+,0}+{\cal M}_{-,0}^*&{\cal M}_{-,0}+{\cal M}_{-,+}^*
\end{array}\right).\quad\phd
\eea

\noi Note that $(\widehat{P}_-+\delta\widehat{P}_-)^{\dag}=\widehat{P}_++\delta\widehat{P}_+$. The coupling to the radiation field now becomes 
$-\big[\big(\widehat{P}_++\delta\widehat{P}_+\big){\cal E}_-+\big(\widehat{P}_-+\delta\widehat{P}_-\big){\cal E}_+\big]$.

\subsection{Conductivity tensor}

The $C_3$ symmetry breaking terms also modify the conductivity and dielectric tensors. The expression for the conductivity tensor in the corrected basis,
$\big|\tilde{\lambda}\big>$, reads:
\bea
\tilde{\sigma}_{ss'}(\omega)
\simeq
\sum_{\lambda,\lambda'}^{0,\pm} 
\frac{i\tilde{\omega}_{\lambda,\lambda'}^2\big<\tilde{\lambda}\big|\big(\widehat{P}_{s}+\delta\widehat{P}_{s}\big)\big|\tilde{\lambda}'\big>
\big<\tilde{\lambda}'\big|\big(\widehat{P}_{s'}+\delta\widehat{P}_{s'}\big)
\big|\tilde{\lambda}\big>\big[f(\widetilde{E}_{\lambda})-f(\widetilde{E}_{\lambda'})\big]}{v\hbar\omega\left(\tilde{\omega}_{\lambda,\lambda'}+\omega+i\eta\right)}\,,
\label{eq:conductivityCorrected}
\eea

\noi where $\widetilde{E}_{\lambda}=E_{\lambda}(t\rightarrow \tilde{t})$ and $\tilde{\omega}_{\lambda,\lambda'}=\omega_{\lambda,\lambda}(t\rightarrow \tilde{t})$. We obtain
the modified conductivity tensor elements:
\bea
\frac{1}{2}\left(\begin{array}{c}\tilde{\sigma}_{xx}(\omega)+\tilde{\sigma}_{yy}(\omega)\\\tilde{\sigma}_{xy}(\omega)-\tilde{\sigma}_{yx}(\omega)\end{array}\right)
&\simeq&(1+2u_x/3)\left(\begin{array}{c}\sigma(\omega)\\\sigma_H(\omega)\end{array}\right)\,,\\
\frac{1}{2}\left(\begin{array}{c}\tilde{\sigma}_{xx}(\omega)-\tilde{\sigma}_{yy}(\omega)\\\tilde{\sigma}_{xy}(\omega)+\tilde{\sigma}_{yx}(\omega)\end{array}\right)
&\simeq&\frac{2}{3}\sigma(\omega)\left(\begin{array}{c}u_x\\u_y\end{array}\right)+
\sum_{\lambda=\pm}\sigma^{\lambda}(\omega)
\left\{
\frac{a}{2t}\frac{\cos(2\pi/3+\lambda\phi)}{\sin(\lambda\phi)\sin(2\pi/3-\lambda\phi)}\left(\begin{array}{c}\phantom{-}eE_x+\zeta u_x\\-eE_y+\zeta u_y\end{array}\right)
\right.\no\\
&&\left.+\frac{\kappa a}{\sqrt{3}t}\big[\cot(2\pi/3-\lambda\phi)+\cot(\lambda\phi)\big]\left(\begin{array}{c}u_x\\u_y\end{array}\right)
\right\}\,,
\eea

\noi where the conductivites appearing on the right hand sides of the above equations are supposed to be calculated for $t\rightarrow \tilde{t}$ and then only linear orders in
the external fields should be kept for the final expression. Note that the correction in $\sigma_{xx}(\omega)+\sigma_{yy}(\omega)$ and
$\sigma_{xy}(\omega)-\sigma_{yx}(\omega)$ originates from the modification of the polarization operator. In contrast, the last two conductivity elements,
$\sigma_{xx}(\omega)-\sigma_{yy}(\omega)$ and $\sigma_{xy}(\omega)+\sigma_{yx}(\omega)$, arise from the modification of the eigenstates.

\subsection{Dielectric tensor and electromagnetic wave propagation}

When the $C_3$ breaking terms are present, the dielectric tensor is diagonalized for electric fields which are admixtures of the left and right CP polarizations. The
dielectric functions for the mixed CPs read
\bea
\tilde{\varepsilon}_{\pm}(\omega)=\left(1+\frac{2u_x}{3}\right)\left[\varepsilon(\omega)-1\right]+1\pm
\left(1+\frac{2u_x}{3}\right)\varepsilon_H(\omega)\sqrt{1
-\frac{\left[\tilde{\sigma}_{xy}(\omega)+\tilde{\sigma}_{yx}(\omega)\right]^2+\left[\tilde{\sigma}_{xx}(\omega)-\tilde{\sigma}_{yy}(\omega)\right]^2}
{\left[2\left(1+2u_x/3\right)\sigma_H(\omega)\right]^2}}\,.
\eea

\noi In the $C_3$ symmetric case, we found that for $\omega\gg\omega_{\pm,0}$, $\sigma(\omega)\simeq i{\rm Im}\ph\sigma(\omega)$ and $\sigma_H(\omega)\simeq {\rm
Re}\ph\sigma_H(\omega)$. Thus there is also here a frequency $\tilde{\Omega}$ for which $\tilde{\varepsilon}_+(\tilde{\Omega})=1$. However, here the solution is not purely
right handed, but it has an admixture of a left handed component, i.e. $\widetilde{{\cal E}}_+(\omega)=\cos(\chi/2){\cal E}_+(\omega)-\sin(\chi/2){\cal E}_-(\omega)$. The
admixture is quite small for weak external fields and is reflected in the admixture angle, $\chi$, which for $\omega\gg\omega_{\pm,0}$ is defined as
\bea
\tan\chi=\frac{\sqrt{\left[{\rm Im}\ph\tilde{\sigma}_{xy}(\omega)+{\rm Im}\ph\tilde{\sigma}_{yx}(\omega)\right]^2+
\left[{\rm Im}\ph\tilde{\sigma}_{xx}(\omega)-{\rm Im}\ph\tilde{\sigma}_{yy}(\omega)\right]^2}}{2{\rm Re}\ph\sigma_H(\omega)}\,.
\eea

\noi For weak perturbations $\tan\chi\simeq\chi$ and $\widetilde{{\cal E}}_+(\omega)\simeq{\cal E}_+(\omega)-\chi{\cal E}_-(\omega)/2$. For estimating the mixing 
angle we will consider, $\omega=\Omega$, which defines the onset of strong birefringence for the $C_3$ symmetric case. In this case, we have
\bea
\chi\simeq\frac{2}{3}\left[1+\frac{(d/2)^4}{1+(d/2)^2}\sum_{\lambda=\pm}\omega_{\lambda,0}^3\frac{\cot(2\pi/3-\lambda\phi)+\cot(\lambda\phi)}
{\sqrt{3}(\omega_{-,0}^3+\omega_{+,0}^3)}\right]u+
\sum_{\lambda=\pm}\frac{\omega_{\lambda,0}^3}{\omega_{-,0}^3+\omega_{+,0}^3}\frac{\cos(2\pi/3+\lambda\phi)}{\sin(\lambda\phi)\sin(2\pi/3-\lambda\phi)}\frac{eaE}{2t}\,,
\eea

\noi with $u=|\bm{u}|$ and the dimensionless $ E=|\bm{E}|$m/kV. For the case $\nu=1/6$ and $d=4$, which was discussed in the main text, we set $\omega=\Omega$ and obtain
$\chi\simeq -2.4u+0.4E$. For a dot displacement of $7.5\%$, i.e. $u=0.075$ corresponding to 8.7nm, or an electric field $E=0.45$ we obtain $\chi\simeq 0.18$ leading to
$\chi/2=9\%$ admixture of CPs.

\end{widetext}

\end{document}